\documentclass[12pt]{article}
\usepackage{html}
\usepackage{epsf}
\usepackage{psfig}
\begin{document}
\title{MATTERS OF GRAVITY, The newsletter of the APS Topical Group on 
Gravitation}
\begin{center}
{ \Large {\bf MATTERS OF GRAVITY}}\\ 
\bigskip
\hrule
\medskip
{The newsletter of the Topical Group on Gravitation of the American Physical 
Society}\\
\medskip
{\bf Number 30 \hfill Fall 2007}
\end{center}
\begin{flushleft}
\tableofcontents
\vfill\eject
\section*{\noindent  Editor\hfill}
David Garfinkle\\
\smallskip
Department of Physics
Oakland University
Rochester, MI 48309\\
Phone: (248) 370-3411\\
Internet: 
\htmladdnormallink{\protect {\tt{garfinkl-at-oakland.edu}}}
{mailto:garfinkl@oakland.edu}\\
WWW: \htmladdnormallink
{\protect {\tt{http://www.oakland.edu/physics/physics\textunderscore people/faculty/Garfinkle.htm}}}
{http://www.oakland.edu/physics/physics_people/faculty/Garfinkle.htm}\\

\section*{\noindent  Associate Editor\hfill}
Greg Comer\\
\smallskip
Department of Physics and Center for Fluids at All Scales,\\
St. Louis University,
St. Louis, MO 63103\\
Phone: (314) 977-8432\\
Internet:
\htmladdnormallink{\protect {\tt{comergl-at-slu.edu}}}
{mailto:comergl@slu.edu}\\
WWW: \htmladdnormallink{\protect {\tt{http://www.slu.edu/colleges/AS/physics/profs/comer.html}}}
{http://www.slu.edu//colleges/AS/physics/profs/comer.html}\\
\bigskip
\hfill ISSN: 1527-3431

\begin{rawhtml}
<P>
<BR><HR><P>
\end{rawhtml}
\end{flushleft}
\pagebreak
\section*{Editorial}

The next newsletter is due February 1st.  This and all subsequent
issues will be available on the web at
\htmladdnormallink 
{\protect {\tt {http://www.oakland.edu/physics/Gravity.htm}}}
{http://www.oakland.edu/physics/Gravity.htm} 
All issues before number {\bf 28} are available at
\htmladdnormallink {\protect {\tt {http://www.phys.lsu.edu/mog}}}
{http://www.phys.lsu.edu/mog}

Any ideas for topics
that should be covered by the newsletter, should be emailed to me, or 
Greg Comer, or
the relevant correspondent.  Any comments/questions/complaints
about the newsletter should be emailed to me.

A hardcopy of the newsletter is distributed free of charge to the
members of the APS Topical Group on Gravitation upon request (the
default distribution form is via the web) to the secretary of the
Topical Group.  It is considered a lack of etiquette to ask me to mail
you hard copies of the newsletter unless you have exhausted all your
resources to get your copy otherwise.

\hfill David Garfinkle 

\bigbreak

\vspace{-0.8cm}
\parskip=0pt
\section*{Correspondents of Matters of Gravity}
\begin{itemize}
\setlength{\itemsep}{-5pt}
\setlength{\parsep}{0pt}
\item John Friedman and Kip Thorne: Relativistic Astrophysics,
\item Bei-Lok Hu: Quantum Cosmology and Related Topics
\item Gary Horowitz: Interface with Mathematical High Energy Physics and
String Theory
\item Beverly Berger: News from NSF
\item Richard Matzner: Numerical Relativity
\item Abhay Ashtekar and Ted Newman: Mathematical Relativity
\item Bernie Schutz: News From Europe
\item Lee Smolin: Quantum Gravity
\item Cliff Will: Confrontation of Theory with Experiment
\item Peter Bender: Space Experiments
\item Jens Gundlach: Laboratory Experiments
\item Warren Johnson: Resonant Mass Gravitational Wave Detectors
\item David Shoemaker: LIGO Project
\item Stan Whitcomb: Gravitational Wave detection
\item Peter Saulson and Jorge Pullin: former editors, correspondents at large.
\end{itemize}
\section*{Topical Group in Gravitation (GGR) Authorities}
Chair: Dieter Brill; Chair-Elect: 
David Garfinkle; Vice-Chair: Stan Whitcomb. 
Secretary-Treasurer: Vern Sandberg; Past Chair:  \'{E}anna Flanagan;
Delegates:
Vicky Kalogera, Steve Penn,
Alessandra Buonanno, Bob Wagoner,
Lee Lindblom, Eric Poisson.
\parskip=10pt

\vfill
\eject

\section*{\centerline
{we hear that \dots}}
\addtocontents{toc}{\protect\medskip}
\addtocontents{toc}{\bf GGR News:}
\addcontentsline{toc}{subsubsection}{
\it we hear that \dots , by David Garfinkle}
\parskip=3pt
\begin{center}
David Garfinkle, Oakland University
\htmladdnormallink{garfinkl-at-oakland.edu}
{mailto:garfinkl@oakland.edu}
\end{center}

Clifford Will has been elected as a member of the National Academy of Sciences

Jim Wilson has won the APS Bethe prize

Martin Bojowald and Thomas Thiemann have won the  Xanthopoulos prize

Yoichi Aso has won the GWIC thesis prize

Stan Whitcomb has been elected Vice-chair of GGR

Lee Lindblom and Eric Poisson have been elected Members at large of
the Executive Committee of GGR

Hearty Congratulations!
\vfill\eject

\section*{\centerline
{News from the} \centerline{International Society on General Relativity and Gravitation}}
\addtocontents{toc}{\protect\medskip}
\addcontentsline{toc}{subsubsection}{
\it News from the GRG Society, by Abhay Ashtekar}
\parskip=3pt
\begin{center}
Abhay Ashtekar, Pennsylvania State University
\htmladdnormallink{ashtekar-at-gravity.psu.edu}
{mailto:ashtekar@gravity.psu.edu}
\end{center}

{\centerline{\bf Near-term Goals and Plans}}
\bigskip

nternational Society on GRG was formed in 1971 as the successor
to the International Committee on GRG. It is Affiliated Commission 2
of the International Union of Pure and Applied Physics (IUPAP), and
within IUPAP is one of the participants in its Particle, Nuclear
Astrophysics and Gravitation International Committee (PANAGIC).

The Society has undergone impressive growth in membership and, more
importantly, in its intellectual reach. Initially, its primary focus
was on analytical general relativity. Now its research areas also
include a large number of other fields: geometric analysis,
numerical relativity, gravitational wave physics and associated
instrumentation and data analysis, relativistic astrophysics,
physical cosmology, early universe, quantum cosmology, quantum
geometry, quantum gravity and string theory. It is even more
impressive that significant advances continue to occur on all these
diverse frontiers. In many cases, some of the central challenges
that were posed by the pioneers some 20 years ago have been
addressed. Here are just a few examples. LIGO has achieved the
desired sensitivity along the entire range of its frequency band.
Stable binary black hole simulations are now feasible and capable of
providing interesting astrophysical insights as well as necessary
templates for data analysis. Global existence theorems for small
initial data have emerged in full non-linear general relativity.
The mathematical status of quantum field theory in curved space-times
has been elevated to that of quantum field theory in flat
space-time. Detailed analyses have emerged indicating how
non-perturbative features of quantum gravity can lead to the
resolution of the big-bang singularity and of the information loss
quandary. In all cases, rather than just closing doors, successes
have opened new avenues to address challenges at the next level.
This is why our field continues to attract so many of the best and
the brightest of young researchers.

A priority for the International Society for General Relativity and
Gravitation is to ensure that this dynamic growth does not come at
the cost of fragmentation of our discipline. The tri-annual
conferences of the Society fulfill an important need in this
respect. The Society will continue to preserve their intellectual
diversity by incorporating new frontiers on an ongoing basis. GR18
at Sydney, for example, was a joint conference with Amaldi7. The
International Committee of the Society will strive to use this
momentum to continue dialogues and cross-fertilizations of ideas
between different areas of our diverse field. We hope to see a
significant growth in the number of participants in GR19 which is
scheduled to take place in Mexico City in July 2010.

As most of you know, the publication organ of the Society, the
journal {\it General Relativity and Gravitation,} was recently
revitalized with a new Editorial Board, representing the current
intellectual richness of our field. Under the joint leadership of
Professors George Ellis and Hermann Nicolai, the journal has made a
significant leap in just over a year. During its meeting at Sydney,
the International Committee proposed numerous new initiatives to
enhance the number of topical reviews and special issues dedicated,
for example, to proceedings of focused workshops. In this endeavor,
we look forward to an active involvement of the community.

Over the next three years, the International Society will strengthen
its ties with the National Societies in our field. We understand
that funding agencies in some of the developing countries
still have considerable flexibility. The International Society will
provide active help in communicating the intellectual excitement and
the growing significance of our field to the appropriate agencies,
world-wide, in the hope of attracting resources that our growing
field richly deserves. We also hope to enhance international
cooperation between National Societies, facilitating regional
meetings that transcend national boundaries. Regular mailings sent
out by Professor MacCallum through the Society's  `hyperspace' service
already keeps the international community aware of various events
and opportunities in our field. This service will soon be enhanced.

While the membership of the Society has reached a new high, by the 
standards of International Societies it is still quite low. Since a
large number of researchers who actively publish in our field are
yet to become members, the International Committee believes that
membership can be significantly increased in the coming years. We
seek your help in persuading your friends and colleagues, not only
in general relativity but also geometric analysis, high energy
physics, cosmology, astrophysics and {\it especially experimental
gravitational physics}, to become members. For information on how to
join, please contact Ms Randi Neshteruk
at
\htmladdnormallink{rxh1-at-psu.edu}
{mailto:rxh1@psu.edu}
or visit 
\htmladdnormallink{www.maths.qmul.ac.uk/grgsoc/}{http://www.maths.qmul.ac.uk/grgsoc/}

The current US representatives on the International committee are
Professors Gary Horowitz (UC, Santa Barbara), James Isenberg (U of
Oregon) and Jorge Pullin (LSU). The Chairperson of the local
organizing committee for GR19 is Professor Hernando Quevedo (UNAM)
and of the Scientific organizing committee is Professor Donald
Marolf (UC, Santa Barbara). The Executive Committee of the Society
consists of Professors Malcolm MacCallum (QMUL, London; Secretary),
Clifford Will (Washington U; Deputy President) and Abhay Ashtekar
(Penn State; President). Please do not hesitate to contact any of us
with suggestions or questions.

\bigskip

\centerline{\bf International Prizes and Awards}

\bigskip
The Society administers several international prizes and awards and
plans are underway to enhance these recognitions, especially for
younger researchers in our field.  The following awards were given
during (or soon after) the GR18/Amaldi7 conference in Sydney.

{\sl The International Xanthopoulos Prize:} This prize was
instituted by the FORTH Foundation of Greece to honor the memory
of Basilis Xanthopoulos, a young relativist whose prolific and
most promising career was brought to an abrupt end, while giving a
seminar, by a deranged assassin. It is given tri-annually to
researchers who are below the age of 40 or who have had no more
than 12 years of research experience following their PhD, and have
made outstanding (preferably theoretical) contributions to
gravitational physics. It carries with it a certificate,
travelling expenses to the GRG conferences and a cash award of
\$10,000. The sixth Basilis Xanthopoulos Prize is awarded jointly
to {\it Martin Bojowald} and {\it Thomas Thiemann} for their
seminal and complementary contributions to the development of
background-independent quantum gravity. The citations were as
follows:

{\narrower\smallskip\noindent{Thomas Thiemann has made major and
highly original advances in the mathematical foundations and
formulation of loop quantum gravity, including the discovery of
what are now being called the Thiemann Identities and the
construction of coherent states, both of which have advanced the
program of consistently interpreting the Hamiltonian constraint
and connecting loop quantum gravity with the classical Einstein
equations. This work has been a major stimulus to the study of
background-independent quantum gravity.\smallskip}}

{\narrower\smallskip\noindent {Martin Bojowald has made deeply
original progress on linking quantized gravity with the classical
equations and concrete physical phenomena. By showing how notions
of symmetry may be incorporated into loop quantum gravity,
Bojowald opened a new approach to quantum cosmology. His
``effective equations", which provide a semi-classical approach to
loop quantum cosmology, have already led to striking results on
the avoidance of a cosmological singularity and on early
inflation. This work has stimulated a great deal of new work on
quantum cosmology.\smallskip}}

\medskip

{\sl The Gravitational Wave International Committee {\rm (GWIC)}
Thesis Prize:} This annual prize evolved from the earlier bi-annual
LIGO thesis prize. It now includes all the gravitational wave
projects world-wide and is thus an international honor awarded by
GWIC, the Society serving as trustees for the funds. It carries a
certificate and a cash award of \$1,000. The first GWIC Thesis Prize
was presented to {\it Yoichi Aso} for his thesis ``Active Vibration
Isolation for a Laser Interferometric Gravitational Wave Detector
using a Suspension Point Interferometer'', submitted to the
Department of Physics, Faculty of Science, University of Tokyo. His
research was carried out under the supervision of Professor Kimio
Tsubono.

The criteria for the award are: 1) originality and creativity of
the research, 2) its importance to the field of gravitational
waves and gravitational wave detection, broadly interpreted, and,
3) how it supports GWIC's goals to promote construction and
exploitation of gravitational-wave detectors, to foster
development of new or enhanced gravitational-wave detectors, and
to support the development of gravitational-wave detection as an
astronomical tool generally.

\medskip

{\sl James B. Hartle Awards:} These international awards are made to
students making the best presentations at the tri-annual GRG
conferences. For the GR18/Amaldi7 conference, each award carried a 3
year membership of the Society and a cash prize of \$50.

Nine students won this honor: {\sl Celine Cattoen} (Wellington;
Session on Dark Energy and the Cosmological Constant); {\sl Ertan
Goklu} (Bremen; Session on Other Quantum Aspects); {\sl Andrew
Moylan} (ANU; Session on CMB, Large Scale Structure \& Gravitational
Lenses); {\sl Jennifer Seiler} (AEI; Session on Numerical
Relativity); {\sl Hanns Selig} (Bremen; Session on Other
Experimental \& Observational Tests of Gravitational Theories); {\sl
Victor Taveras} (Penn State; Session on Quantum Aspects of Black
Holes); {\sl Robert Ward} (Caltech; Session on R\&D for Advanced
Ground Based GW Detectors); {\sl Lila Warszawski} (Melbourne;
Session on Relativistic Astrophysics); and, {\sl Shuichiro Yokoyama}
(Kyoto; Session on the Early Universe and Pre-Big Bang).

\vfill\eject

\section*{\centerline
{LIGO/GEO/Virgo begin working together}}
\addtocontents{toc}{\protect\medskip}
\addcontentsline{toc}{subsubsection}{
\it LIGO/GEO/Virgo work together, by Peter Saulson}
\parskip=3pt
\begin{center}
Peter Saulson, Syracuse University
\htmladdnormallink{saulson-at-physics.syr.edu}
{mailto:saulson@physics.syr.edu}
\end{center}

   The past year has seen outstanding progress in the global search for gravitational waves. The three LIGO interferometers have been carrying out nearly continuous observations at their design sensitivity for the 5th Science Run (S5), joined for substantial periods by the GEO 600 interferometer. S5’s goal of one year’s worth of coincident data will soon be achieved. Meanwhile, discussions were held between the LIGO Laboratory and the LIGO Scientific Collaboration (which includes the members of GEO) and Virgo and EGO (the European Gravitational Observatory which operates the Virgo interferometer) on arrangements for joint observing and data analysis. Those discussions came to a successful conclusion, as marked by two events in 2007: the signing of a Memorandum of Understanding laying out the agreement to observe and analyze data together, and then, on 
May 18, 2007, the start of the first Virgo Science Run (VSR1) in coordination with LIGO’s S5.

   The agreement serves as a model for global network analysis of data from all gravitational wave detectors, as they come on line. The basic structure is not a merger, but instead a commitment from independent projects to work together on both the operation of their detectors and on data analysis. The collaboration on data analysis is especially close. Except for closing out of papers on pre-agreement observations, all analysis will be carried out by joint teams, and will be published in papers bearing the names of the members of all of the projects. 

   To make this agreement a reality, members from the various projects have begun working together through a variety of structures. Data analysts from the LSC and from Virgo now work together in joint Data Analysis Groups (one for each of the major signal categories: bursts, inspirals, pulsars, and stochastic background.) These data analysis groups will have joint review committees attached to them to vet observational results. On the operational side of things, a joint Run Planning Committee coordinates plans for observing, maintenance, commissioning, and upgrades. Collaboration meetings are now joint meetings between the two collaborations. Two will be held each year in Europe, and two in the U.S. The inaugural joint meeting was held in Baton Rouge on 
March 19-22; followed by another in Pisa on 
May 21-24, coordinated with the start of VSR1. A subsequent joint meeting was 
held on July 23-26 at MIT, and another will be held in Hannover on  
October 22-25.

As important as this present work is, perhaps even more important are the plans for the future. The science runs S5 and VSR1 will finish this fall. Then, LIGO and Virgo will each undertake a program of incremental upgrades. LIGO, for example, will upgrade two of its interferometers (the 4-km interferometers at the Hanford and Livingston sites) with new lasers and readout optics. These Enhanced LIGO interferometers are hoped to achieve a sensitivity twice as great as they had in their original state. Virgo plans to upgrade its 3-km interferometer to become Virgo+, with similar sensitivity to Enhanced LIGO. While those upgrades are taking place, GEO will operate the GEO 600 interferometer, and LIGO will operate the 2-km interferometer at Hanford with as high a duty factor as it can, consistent with the upgrade work on the 4-km interferometer. TAMA, in Japan, also plans to be on the air; discussions are under way to bring it under the umbrella of the LIGO/Virgo agreement. In addition, the bar detectors AURIGA and those operated by the Rome Group (EXPLORER  and NAUTILUS) will be collecting data; discussions are also under way with those groups. 

It is hoped that this upcoming upgrade phase will be completed in less than two years, to be followed by a substantial (of order 1 year) science run of the upgraded interferometers. Time pressure comes from the happy prospect of the start of construction of Advanced LIGO in the U.S. (and, in parallel, Advanced Virgo with comparable sensitivity.) Advanced LIGO is designed to have sensitivity ten (or more) times greater than initial LIGO. At that sensitivity, it is expected that signals will be detected regularly (once/month or perhaps even more frequently.) Those detectors, to come on line around 2015, will inaugurate the era of gravitational wave astronomy. On that time scale, there are also good prospects for detectors of similar sensitivity in Japan (LCGT) and Australia (AIGO.)

Funding for the start of Advanced LIGO is in the NSF budget bills now making their way through both houses of the U.S. Congress. Assuming successful passage and signing of those bills, Advanced LIGO hardware is expected to be ready for installation at the LIGO observatories at the end of 2010.

\vfill\eject

\section*{\centerline
{GWIC - Ten Years on}}
\addtocontents{toc}{\protect\medskip}
\addcontentsline{toc}{subsubsection}{
\it GWIC - Ten Years on , by Stan Whitcomb}
\parskip=3pt
\begin{center}
Stan Whitcomb, LIGO 
\htmladdnormallink{stan-at-ligo.caltech.edu}
{mailto:stan@ligo.caltech.edu}
\end{center}

The Gravitational Wave International Committee (GWIC) celebrates its tenth birthday this year.  It was formed in 1997 to facilitate international collaboration and cooperation in the construction, operation and use of the major gravitational wave detection facilities world-wide.  GWIC’s goals are broad and far-reaching:
\begin{itemize}
\setlength{\itemsep}{-5pt}
\setlength{\parsep}{0pt}
\item Promote international cooperation in all phases of construction and exploitation of gravitational-wave detectors;
\item Coordinate and support long-range planning for new instrument proposals, or proposals for instrument upgrades;
\item Promote the development of gravitational-wave detection as an astronomical tool, exploiting especially the potential for coincident detection of gravitational-waves and signals from other fields (photons, cosmic-rays, neutrinos);
\item Organize regular, world-inclusive meetings and workshops for the study of problems related to the development or exploitation of new or enhanced gravitational-wave detectors, and foster research and development of new technology;
\item Represent the gravitational-wave detection community internationally, acting as its advocate;
\item Provide a forum for the laboratory directors to regularly meet, discuss, and plan jointly the operations and direction of their laboratories and experimental gravitational-wave physics generally.
\end{itemize}
GWIC derives its formal standing in the international physics community through IUPAP (International Union of Pure and Applied Physics).  IUPAP was established in 1922, with the broad mission to assist in the worldwide development of physics, to foster international cooperation in physics, and to help in the application of physics toward solving problems of concern to humanity.  One of the IUPAP Working Groups (WG.4: Particle and Nuclear Astrophysics and Gravitation International Committee or PaNAGIC) has “adopted” GWIC as a specialized sub-field panel, which is done when PaNAGIC determines that a sub-panel will be useful in promoting convergence of large international projects.  The chairman of GWIC is automatically a member of PaNAGIC.  These ties also give GWIC a formal link to the International Society on General Relativity and Gravitation, which is an Affiliated Commission of IUPAP and a participant in PaNAGIC.  
\vskip0.25truein
{\it Who is GWIC?}
\vskip0.25truein
The membership of GWIC represents all of the world’s active gravitational wave projects, both ground-based and space-based.  Each project has either one or two members on GWIC depending on size.  Because the GWIC representatives are generally the leaders of each project, GWIC has access to the broader expertise throughout the community.  GWIC also includes representation from the International Society on General Relativity and Gravitation and from the astrophysics/theoretical relativity community.  

This year has seen important changes in the leadership of GWIC.  At its meeting in July, GWIC selected Jim Hough (GEO) as its new chair, succeeding Massimo Cerdonio (AURIGA) and before that Barry Barish (LIGO).  Earlier this year, Sam Finn stepped down as Executive Secretary—Sam had held this post since GWIC’s inception, and we owe him a debt of gratitude for his service.
\vskip0.25truein
{\it What does GWIC do?}
\vskip0.25truein
GWIC has some very easily identifiable activities that many of you will recognize:
\begin{itemize}
\setlength{\itemsep}{-5pt}
\setlength{\parsep}{0pt}
\item GWIC sponsors the biennial Edoardo Amaldi Conferences on Gravitational Waves (see the report on Amaldi 7 by Jorge Pullin in this issue of MOG).  The Amaldi meeting is considered by many in the gravitational wave community to be their most important international gathering.  The members of GWIC serve as the Scientific Organizing Committee for the Amaldi meetings.  The next (8th) Amaldi meeting will be held at Columbia University from June 21-26, 2009. 
\item In 2006, GWIC established an international prize, to be awarded annually to an outstanding Ph. D. thesis based on research in gravitational waves.  The 2006 GWIC Thesis Prize was just presented at the Sydney 7th Amaldi meeting to Dr. Yoichi Aso, for his research performed at the University of Tokyo.  A first Announcement of the 2007 GWIC Thesis Prize, to be presented at the LISA Symposium in Barcelona in June 2008, is attached at the end of this article.
\end{itemize}
However, I would argue that GWIC’s most important contributions are less concrete.  By bringing together the leaders of the different projects on a regular basis, it has helped break-down the barriers and improved communication among the various gravitational wave projects.  The growing collaboration among the various gravitational wave projects has been triggered in large part by discussions which have taken place at GWIC meetings.  In particular, the recent agreement between LIGO/GEO and Virgo to analyze their data together has its roots in the GWIC meeting in the summer of 2005.

Along this line, one of the major up-coming activities for GWIC was commissioned at its July meeting: Jay Marx (LIGO) was appointed chair of a committee to prepare a global road-map for the field of gravitational wave science, with the perspective to optimize the global science in the field.  The charge to the committee is to cover both ground- and space-based detectors with 30-year horizon.  The final report will use broad input from the communities affected to identify relevant science opportunities and the facilities needed to address them.  We hope that this study will help focus the R\&D for the next few years and guide the funding agencies to support the highest priority projects.  

\vfill\eject

{\centerline {\bf GWIC Thesis Prize}}

{\centerline {\bf First Announcement of 2007 Prize}}
\vskip0.5truein
The Gravitational Wave International Committee (GWIC) was formed to promote international collaboration and cooperation in the construction, operation and use of gravitational wave detection facilities world-wide.  To this end, GWIC has established an annual prize for the outstanding Ph.D. thesis based on research in gravitational waves.

Members of the broader gravitational wave community are invited to nominate students who have performed notable research on any aspect of gravitational waves science. Theses will be judged on 1) originality and creativity of the research, 2) importance to the field of gravitational waves and gravitational wave detection, broadly interpreted, and 3) clarity of presentation in the thesis.  

{\bf Eligibility:} The award is made on a calendar year basis.  Theses must have been accepted by their institutions between January 1, 2007 and December 31, 2007 to qualify for consideration.  It is expected that many of the nominations will come from the member projects of GWIC, but this is not a requirement.

A committee representing the GWIC member projects will evaluate the nominations and select the winner.  Nominated theses may be in any language -- the selection committee will use consultants to help evaluate theses if they do not possess the required linguistic breadth. The selection committee will make the final determinations about eligibility.

The GWIC Thesis Prize will be presented at the LISA Symposium in Barcelona, Spain, 16-20 June 2008. The recipient will receive a certificate of recognition and a prize of \$1,000.

{\bf Nominations:} A Call for Nominations will be issued approximately November 1, 2007 with instructions about how to submit a nomination.  
 

\vfill\eject

\section*{\centerline
{Quasi-local definitions of energy in general relativity}}
\addtocontents{toc}{\protect\medskip}
\addtocontents{toc}{\bf Research Briefs:}
\addcontentsline{toc}{subsubsection}{
\it Quasi-local energy, by Bjoern S. Schmekel}
\parskip=3pt
\begin{center}
Bjoern S. Schmekel, The University of California, Berkeley 
\htmladdnormallink{schmekel-at-berkeley.edu}
{mailto:schmekel@berkeley.edu}
\end{center}

Defining energy is a surprisingly difficult problem in general relativity. 
For instance, the energy density of the gravitational field of a planet at 
a particular point could be determined by a comoving observer measuring the 
kinetic energy of a freely falling object. Due to the equivalence  
principle, both the object and the observer fall at equal rates. Therefore, 
the observer would
not assign any energy to the object. Other observers like an observer 
who is at rest with
respect to the planet would measure different values. This raises the 
question of how energy 
depends on the choice of an observer which violates the philosophy of 
general relativity
whose tensorial equations are independent of the used reference system. 

In classical electrodynamics the stress-energy tensor is a measure of the 
energy 
and momentum transported
by the electromagnetic field due to a source distribution $j^\mu$. A similar
construction in general relativity leads to the so-called Bel-Robinson tensor 
$T_{\mu \nu \rho \sigma}$ \cite{Bel:1958,Robinson:1958,Bel:1959} which can 
be thought of as being
induced by a stress-energy tensor $T_{\mu \nu}$.
Its physical meaning however remains unknown since it does not even have
units of energy density. This is a consequence of the equivalence
principle which equates the gravitational mass (the "charge" of gravity)
with the inertial mass. The source term, i.e. $j_\mu$ in electrodynamics
and $T_{\mu \nu}$ in general relativity, does not contain the energy of the
gravitational field. However, since the equations of general relativity 
are non-linear
there may be a non-linear contribution to the stress-energy.
For instance, gravitational waves do not pass through each other without 
distortion.

Due to the absence of Stokes theorem for second ranked tensors conserved 
quantities
do not exist.

Landau and Lifshitz were able to prove that the stress-energy-momentum 
pseudotensor
\begin{eqnarray}
16 \pi G t^{\mu \nu} = (-g)^{-1} \left [ (-g) \left ( g^{\mu \nu} g^{\rho \sigma}
- g^{\mu \rho} g^{\nu \sigma} \right ) \right ]_{,\rho \sigma} - 2 G^{\mu \nu}
\end{eqnarray}
is the only symmetric pseudotensor constructed only from the metric
such that the four-divergence of the total stress energy vanishes like
$\left [ (-g)(T^{\mu \nu} + t^{\mu \nu} ) \right ]_{,\mu}=0$
and which also vanishes locally in an inertial frame.
The latter requirement is dictated by the equivalence principle as was mentioned
above. However, $t^{\mu \nu}$ still does not transform as a tensor.

Because of the problems associated with defining a local energy density
it may be easier to make sense of the energy enclosed by a boundary. For regions
of finite extend we expect non-zero values because in general a coordinate transformation
can make the connection coefficients vanish at only one point.

Therefore, it seems the only sensible way to define energy is by defining energy
itself and not energy density. Of course this may seem ugly because a local covariant
and tensorial formulation depends on densities evaluated at a point and its
infinitesimally small neighborhood (in order to compute derivatives). A point
remains a point under a Lorentz transformation, but needless to say the size
of a finite region depends on the observer, so obviously such an energy
will depend on the chosen coordinate system. 
It therefore may not be surprising
that the first useful notions of energy were defined at infinity, i.e. they
enclosed the whole system (cf. ADM mass \cite{ADM:1962}, 
Bondi mass \cite{Bondi1964}).
Like a point an infinitely large box does not change its size
under a change of observer. 

A successful definition of quasi-local energy (QLE) was given by Brown and York \cite{Brown:1992br}.
A spacelike three-dimensional hypersurface $\Sigma$ is embedded in a four-dimensional spacetime $M$
which satisfies the Einstein field equations. This embedding defines the
"time-direction". Finally, a two-dimensional boundary $B$, which encloses the energy
of the region we are interested in, is embedded in the three-dimensional hypersurface $\Sigma$. 
$\Sigma$ is enclosed by a three-boundary $^3 B$, and their normals are constrained to
be perpendicular to each other. This restriction ensures the time evolution of the system
is consistent with the presence of the fixed boundary $B$.
In classical mechanics the Hamiltonian is given by the variation of the classical action with
respect to the endpoints times minus one. The Brown-York QLE is derived
by considering the change of the classical action under a displacement 
of the initial
and final hypersurfaces and is given by
\begin{eqnarray}
E = \frac{1}{\kappa} \int_B d^2 x \sqrt{\sigma} \left ( k - k_0 \right )
\end{eqnarray}
where $k$ is the trace of the extrinsic curvature of $B$ and 
$\sigma$ is the metric of $B$.
The surface gravity is denoted by $\kappa$, and $k_0$ is a reference term 
which sets the energy of flat space to zero. The subtraction term has
been criticized. However, it should be mentioned that the ADM energy makes
reference to flat space as well by using ordinary non-covariant derivatives.

For a Schwarzschild black hole the action can be expressed in terms of the
QLE as follows
\begin{eqnarray}
S = \frac{8\pi}{\kappa} \int (N dt) (r f)
\end{eqnarray}
where $-r f$ is the unreferenced QLE. The metric has been
expressed as $ds^2 = - N^2(r) dt^2 + f^{-2}(r) + r^2 d \Omega^2$
where $N(r)=f(r) = \sqrt{1- 2m/r}$. Because the geodesics of infalling 
objects can be determined both inside and outside of a black hole it 
should be possible to assign a value to the energy of the gravitational 
field in both regions. The definition of $E$ can be continued into 
the region inside the horizon \cite{Lundgren:2006fu}. Since both 
$N$ and $f$ become imaginary
inside the horizon $-rf$ needs to be multiplied by $i$ in order to
become real. 

Whether the quantity defined above is useful depends on its properties and
whether applications exist. At infinity the QLE is equal to the ADM energy.
Furthermore, it reduces to the Newtonian binding energy in the 
non-relativistic
limit. In the thermodynamics of black holes the QLE is just the total energy.
Blau and Rollier have shown that the extended Brown-York energy describes
the effective potential of a particle falling into a black 
hole \cite{Blau:2007wj}.
Also worth mentioning is the small sphere limit \cite{Brown:1998bt}. 
In this limit
the QLE reduces to the energy of the enclosed matter. The gravitational
binding energy only shows up in higher orders of the radius which emphasizes
the fact that one cannot make sense of the energy of the gravitational 
field locally.
This may be seen as a hint that point particles do not 
exist \cite{Lundgren:2006fu}.

The most serious drawback is that fact that not all physically interesting
boundaries $B$ can be embedded in a reference space which is typically
taken to be $\mathcal{R}^3$ leading to a non-existence of the reference term. 
An important example is the horizon of a Kerr black hole. 
Usually only energy differences are important, so the absence of the
reference term might not lead to problems. However, the stability of
flat space rests on the fact that every non-flat spacetime contains more
energy that the flat ground state \cite{Schon:1981vd}. Therefore, 
Epp \cite{Epp:2000zr}
and subsequently Liu und Yau \cite{Liu:2003bx} considered a modification of
the Brown-York energy which does not need the three-boundary
$^3 B$. Rather, the two-boundary $B$ is embedded directly
into the four-dimensional spacetime $M$, and the orthogonality condition
is not applicable anymore. Such an embedding does always exist.
However, it is not unique. The absence of the orthogonality condition
implies that the observer is not at rest anymore with respect to $B$.
Denoting the trace of the normal momentum surface density by $l$ which measures
the expansion of $B$ in time the boost-invariant QLE becomes
\begin{eqnarray}
E = \frac{1}{\kappa} \int_B d^2 x \sqrt{\sigma} \left ( \sqrt{k^2-l^2} - \sqrt{k_0^2-l_0^2} \right )
\end{eqnarray}
The most attractive feature of the boost-invariant QLE is certainly
its independence of the observer. It can attain complex values if
$B$ is located inside the event horizon of a black hole. Also, the
integrand of the unreferenced boost-invariant QLE is always positive (if real),
whereas the Brown-York QLE depends on the extrinsic curvature
of $B$ which can be positive or negative. 

Ultimately, the usefulness of the described quantities will depend on the
availability of applications. A more thorough review of the problems
associated with quasi-local energy can be found in \cite{lrr-2004-4}.

\vfill\eject

\section*{\centerline
{The current status of cosmic strings}}
\addtocontents{toc}{\protect\medskip}
\addcontentsline{toc}{subsubsection}{
\it The current status of cosmic strings, by  Patrick Peter}
\parskip=3pt
\begin{center}
Patrick Peter, GR$\varepsilon$CO - Institut d'Astrophysique de Paris - Universit\'e Pierre et Marie Curie
\htmladdnormallink{peter-at-iap.fr}
{mailto:peter@iap.fr}
\end{center}

\def\alt{
\mathrel{\raise.3ex\hbox{$<$}\mkern-14mu\lower0.6ex\hbox{$\sim$}}
}

\def\etal{{\it et.al.~}}

The interest in cosmic string (CS) research has seen a renewal recently when
it was observed that fundamental (super)strings could act as actual topological
defects, although with possibly rather different properties. It was even said,
obviously  in an excess of enthusiasm, that the observation of a single 
cosmic string in the sky would be ``a proof'' of the existence of superstring 
theory!

Most of the original idea dates back to the work of Kibble in 1976 who 
realized that if the Higgs mechanism were to take place in the early stages 
of the Universe's expansion, with a very small Hubble radius (and thus Horizon 
size)---recall it was a time during which inflation had not yet acquired its 
status of a paradigm, as it is now part of the standard model of 
cosmology---then even a simple causality argument led to the existence of 
linear topologically stable configurations, called cosmic strings, that 
should have been formed during the symmetry breaking phase transition. The 
evolution of the thus-formed network of cosmic strings was found to have the 
ability to provide, in the long run, a spectrum of scale invariant 
cosmological perturbations that could then have acted as seeds for large 
scale structure formation. Moreover, the order of magnitude of the 
temperature fluctuations in the CMB is $\Delta T/T \sim G_{\mathrm{N}} 
U$, with $U \sim E_{_{\mathrm{GUT}}}^2$ the energy per unit string length, and 
$G_{\mathrm{N}} = M_{\mathrm{P}}^{-2}$ the Newton constant, i.e., the inverse 
of the Planck mass squared. Based on these assumptions, and provided the 
transition was that expected at the GUT scale ($E_{_{\mathrm{GUT}}} \sim 
10^{16}$GeV), then the theoretical number fits the observation, with no 
fine-tuning involved. General references on these topics are
Vilenkin \& Shellard (2000) and Peter \& Uzan (2008).

CS are predicted in almost any conceivable GUT (Jeannerot \etal 2003), and 
they can be used to impose constraints on the high energy parameters, e.g. 
those stemming from supergravity (Rocher \& Sakellariadou 2004). If endowed 
by a current, which happens to be often the case (although it is very much 
model-dependent), a CS network evolution (Ringeval \etal 2007) ends up 
producing vorton states which overclose the Universe, thereby leading to a 
cosmological catastrophe [see, e.g. Cordero-Cid \etal (2002) and Postma 
\& Hartmann (2007)]. 

Models in which Nambu-Goto strings can form a network and evolve are expected 
to reach scaling, as shown numerically (Fig. 1), in the sense that the energy 
density contained in the network eventually behaves as the inverse of the 
square of the horizon scale. This point was subject to some polemics (see 
Vanchurin, Olum \& Vilenkin (2006) and Martins \& Shellard (2006)) but seems 
now to have been settled by Ringeval \etal (2007).

More precise data have shown, however, that the scale invariant spectrum of
primordial perturbations on large scales is not the whole story, and the recent
WMAP sky observation has revealed that the most important contribution of 
the CMB fluctuations originates with an adiabatic coherent source such as 
predicted by the inflation models (see Lemoine \etal 2007 for a summary on 
inflation and Riazuelo \etal 2000 for CS calculations). In the best case, 
cosmic strings represent a small fraction of the CMB fluctuation. The current 
constraints (which are not only model-dependent, but also quite unclear 
because the actual CS perturbation spectrum is not known with certainty) are 
around 10~\% or less (Wyman \etal 2005). Accordingly, CS research seems to be 
marginalized.

There was a renewal of interest with the realization that, contrary to all 
expectations, fundamental superstrings might actually act as CS, although 
with some differences. Of course, these strings should have, from the outset, 
almost Planckian energy per unit length, thereby producing far too high an 
amplitude in the perturbations. But moreover, Witten (1985), based on 
perturbation arguments, had shown that long fundamental BPS strings in 
heterotic theory were cursed with instabilities and had therefore no chance 
to ever be observed. Non-BPS states were also believed to be unstable.

\begin{figure}[t]
\centerline{
\psfig{file=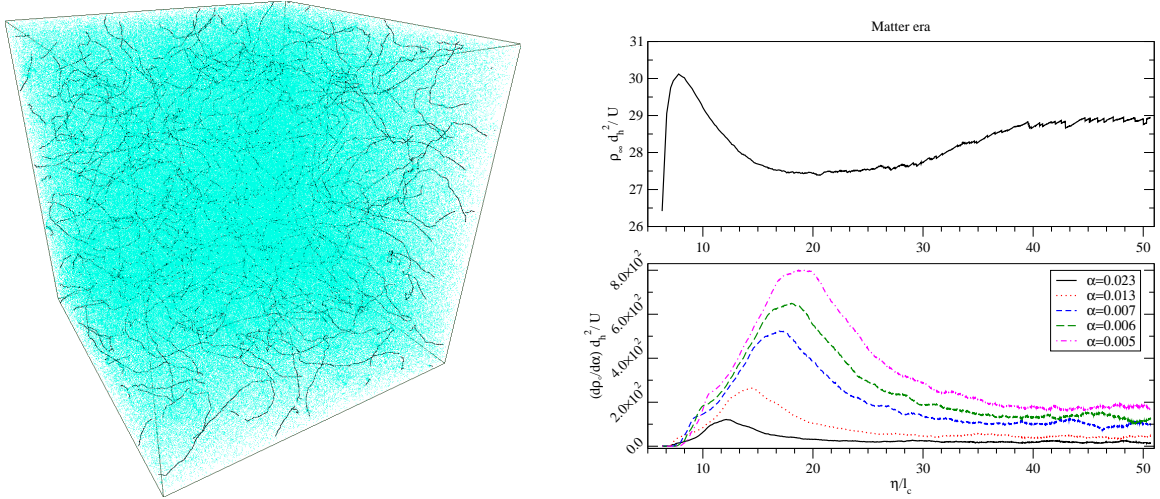,width=2.7in}~~~~~
\psfig{file=enermat.eps,width=3in}}
\caption{Left panel: A typical string loop distribution according to the
latest simulation by Ringeval \etal (2007) during the matter area. The
observable Universe occupies one eighth of the box, whose edge is $100
\ell_{\mathrm{c}}$, with $\ell_{\mathrm{c}}$ being the correlation length of
the Vachaspati-Vilenkin (1984) initial conditions. The right panel shows, for
the same era,  the evolution of the energy density associated with long
strings (top) and loops (bottom) of physical sizes $\ell_{\mathrm{phys}} =
\alpha d_{\mathrm{h}}$, $ d_{\mathrm{h}}$ being the horizon size. The time
variable is the rescaled conformal time $\eta/\ell_{\mathrm{c}}$. These
results show the network reaching scaling, as required.}
\end{figure}

\begin{figure}[h]
\centerline{
\psfig{file=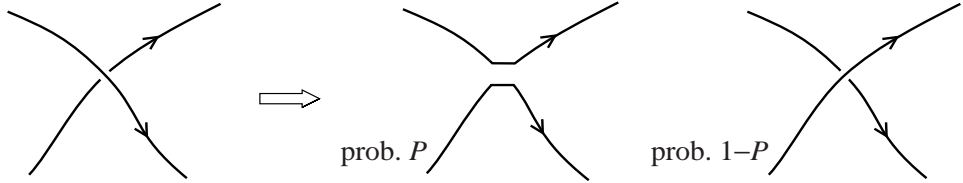,width=5in}}
\caption{String intercommutation}
\end{figure}

All this changed around 2004, when it was found that because of the large 
number of possible geometries for the compact dimensions and with the 
existence of localized branes, the string energy per unit length could be 
much lower than originally thought. Besides, new solutions were found, called 
F, D, and $(p,q)$ strings, that not only resemble the usual local topological 
defects, but also have sufficiently different properties to be discriminated. 
Moreover, those solutions could be stabilized over cosmologically relevant 
time scales. Polchinski (2005) and Majumdar (2005) provide nice overviews of 
these topics.

The most important difference, to date, between fundamental (or cosmological 
size string solutions in supertring theories) and ordinary CS is the 
so-called reconnection probability. Indeed, in a 4D space-time, when two 
topological defects collide, they exchange ends with a probability $P\sim 1$ 
(see Fig. 2). For superstring strings however, because of the 
extra-dimensions in which the strings (or branes) are actually evolving, this 
probability could be drastically reduced $P\ll 1$, leading to a completely 
different network evolution. The same reason implies that the string energy 
per unit length can be much lower than the Planck scale, once the Calabi-Yau 
volume is taken into account. All this leads to observationally compatible 
effects.

This topic is currently seeing some serious development, for these strings 
can easily be embedded in scenarios of brane inflation, where they are 
produced after the inflation epoch. They can produce observable amounts of 
gravitational radiation with particular spectral properties, be involved in 
the (p)reheating mechanism, form structures, help reionization and 
baryogenesis ... in short, the future for research in CS seems still quite 
bright!

{\bf References:}
\hfil\break
A.~Cordero-Cid, X.~Martin and P. Peter, Phys. Rev. {\bf D65}, 083522 (2002) and references
therein.
\hfil\break
R.~Jeannerot, J.~Rocher and M.~Sakellariadou, Phys.Rev. {\bf D68}, 103514 (2003).
\hfil\break
M.~Lemoine, J.~Martin and P.~Peter (Eds.), {\sl Inflationary Cosmology}, Lecture Notes in Physics,
Springer (2007).
\hfil\break
M.~Majumdar, Lecture notes for COSLAB 2004, University of Leiden, Imperial College and
Dhaka University, (2005).
\htmladdnormallink{hep-th/0512062}{http://arxiv.org/abs/hep-th/0512062}
\hfil\break
C.~J.~A.~P.~Martins and E.~P.~S.~Shellard, Phys. Rev. {\bf D73}, 043515 (2006).
\hfil\break
P.~Peter and J.-P.~Uzan, {\sl Primordial Cosmology}, Oxford University Press, to appear (2008).
\hfil\break
J.~Polchinski, AIP Conf.Proc. {\bf 743}, 331 (2005); Int. J. Mod. Phys. {\bf A20}, 3413 (2005).
\hfil\break
M.~Potsma and B.~Hartmann, (2007)
\htmladdnormallink{arXiv:0706.0416}{http://arxiv.org/abs/0706.0416}
\hfil\break
A.~Riazuelo, N.~Deruelle and P.~Peter, Phys. Rev. {\bf D61}, 123505 (2000) and
references therein.
\hfil\break
C.~Ringeval, M.~Sakellariadou, and F.~Bouchet, JCAP {\bf 0702}, 023 (2007).
\hfil\break
T.~Vachaspati and A.~Vilenking, Phys. Rev. {\bf D30}, 2036 (1984).
\hfil\break
V.~Vanchurin, K.~D.Olum, and A.~Vilenkin, Phys. Rev. {\bf D74}, 063527 (2006).
\hfil\break
A.~Vilenkin and E.~P.~S.~Shellard, {\sl Cosmic strings and other topological defects},
Cambridge University Press (2000).
\hfil\break
E.~Witten, Phys. Lett. B{\bf 153}, 243 (1985).
\hfil\break
M.~Wyman, L.~Pogosian, and I.~Wasserman,  Phys.Rev. {\bf D72}, 023513 (2005); Erratum-ibid. {\bf D73}, 089904 (2006).
\hfil\break
\vfill\eject

\section*{\centerline
{Gravitational waves from `mountains' on neutron stars}}
\addtocontents{toc}{\protect\medskip}
\addcontentsline{toc}{subsubsection}{
\it Gravitational waves from `mountains' on neutron stars, by Ian Jones}
\parskip=3pt
\begin{center}
Ian Jones, University of Southampton
\htmladdnormallink{D.I.Jones-at-soton.ac.uk}
{mailto:D.I.Jones@soton.ac.uk}
\end{center}

This article should perhaps begin with a disclaimer: when we talk
about `mountains' on neutron stars we speak not of precipitous
projections from the surface, but rather of large scale mass
asymmetries. Over the last decade a series of papers has appeared in
the literature attempting to calculate just how large such mountains
might be.  There is a strong motivation behind this: In a spinning
star such asymmetries directly source gravitational wave emission, and
with the LIGO and GEO600 detectors up and running, and VIRGO close
behind, accurate modeling of this class of potential source is vital.
As emphasized recently by Owen (2005), having estimates of possible
mountain sizes enables us to assess the significance not just of
gravitational wave detections but (currently more usefully!) of
\emph{upper bounds}, telling us when the physics at least would permit
a detection.

More accurately, if a neutron star spins about some axis $Oz$, we are
concerned with the difference between the $I_{xx}$ and $I_{yy}$ pieces
of its moment of inertia tensor, as a difference between these produce
a characteristic gravitational wave strain of (Abbott et al. 2007):
\begin{equation}
\label{eq:h}
h = \frac{4G}{c^4} \frac{(I_{yy}-I_{xx}) \Omega^2}{r} ,
\end{equation}
where $\Omega$ denotes the star's angular spin frequency and $r$ its
distance from Earth.  The asymmetry $(I_{yy}-I_{xx})$ is often
parameterized in dimensionless form as a so-called \emph{ellipticity}
parameter
\begin{equation}
\label{eq:epsilon}
\epsilon = \frac{I_{yy}-I_{xx}}{I_{zz}} ,
\end{equation}
but it is worthwhile remembering that only the combination $\epsilon
I_{zz} \equiv I_{yy}-I_{xx}$ appears in the formula for $h$. As will
be described below, typical calculations for canonical neutron stars
place $\epsilon$ somewhere around $10^{-6}$ or less, corresponding to
an equatorial radius difference of no more than $1$ cm, so this is
rather a far cry from Mount Everest.  Nevertheless, an ant unlucky
enough to find himself placed on such a star would have to expend
$10^{11}$ ergs to climb up the slope!


If neutron stars consisted of nothing but a ball of self-gravitating
perfect fluid, no such asymmetry could exist.  However, strains in the
solid crust or Lorentz forces connected with the internal magnetic
field can support deformations, and it is to these that theorists have
turned to try and compute possible $\epsilon$ values.  Crustal strains
have attracted more attention, possibly because of the large field
strengths needed to make the magnetic contribution important.

Historically, strains in the solid crust were of interest to early
modelers, who thought that crust fracture might explain the
phenomenon of pulsar glitches.  Calculations soon showed that, for the
Vela pulsar at least, such an explanation wasn't viable---there simply
wasn't enough elastic energy available to account for the frequent
large glitches.  

However, interest in crustal deformations was rekindled in 1998, the
inspiration coming from X-ray physics.  New satellites succeeded in
measuring oscillations in the Low mass X-ray binary (LMXB) systems,
and it seemed the derived spin frequencies were unexpectedly tightly
clustered in the interval $300$--$600$ Hz.  As Bildsten (1998) pointed
out, this was suggestive of the spinning-up stars hitting a
`gravitational wave wall', where the spin-up accretion torque was
balanced by a gravitational wave spin-down torque, the steep spin
frequency dependence of the latter serving to cluster the equilibrium
frequencies.  Such a hypothesis had been advanced previously, first in
outline by Papaloizou \& Pringle (1978), and then in more detail by
Wagoner (1984).  However, Bildsten suggested that mountains (rather
than fluid oscillation modes) might be responsible for the emission,
and even suggested an ingenious way of manufacturing the necessary
$\epsilon \approx 10^{-7} \rightarrow 10^{-8}$ deformations: if
temperature asymmetries were (for whatever reason) intrinsic to the
accretion process, the temperature-dependent nuclear reactions
undergone by the accreting crustal matter would be shifted, with the
resulting density perturbations providing the necessary mass
quadrupole to produce the gravitational wave torque.

This idea was taken up in detail by Ushomirsky, Cutler \& Bildsten
(2000; hereafter UCB), who found that temperature asymmetries at the
$5\%$ level in the inner crust would provide the necessary mountain.
They also calculated a bound on $\epsilon$ that assumed only that
strains in the elastic crust were responsible for the deformation,
making no further assumptions about its cause (e.g. temperature
asymmetries).  This essentially involved balancing the gravitational
and pressure forces (which favor a spherical configuration) against
the shear stress forces in the crust (which cause asymmetry).  Very
roughly, their result was $\epsilon \le 10^{-7} (\sigma_{\rm
  max}/10^{-2})$, where $\sigma_{\rm max}$ is the breaking strain of
neutron star crustal matter.  Unfortunately, this number is very
uncertain; $10^{-5} < \sigma_{\rm max} < 10^{-2}$ for terrestrial
materials, so parameterizing in terms of $\sigma_{\rm max}/10^{-2}$ is
probably rather optimistic.  Nevertheless, for the remainder of this
article we will assume a breaking strain of $10^{-2}$ in all quoted
ellipticities; the results can be scaled to any other breaking strain
in a linear manner.

This maximization problem was developed further by Haskell, Jones \&
Andersson (2006), who extended the treatment in a number of ways.
They considered a variety of stellar models, including models where the
core structure was computed relativistically, stars with both
accreted and non-accreted crusts, and models where the perturbations
in the gravitational potential were retained (these were neglected in
the UCB analysis).  The main conclusion was that there was little
difference between the maximum $\epsilon$ values for the accreted and
non-accreted crust, but that the more accurate treatment of the
gravitational potential, together with an improved treatment of
boundary conditions, resulted in $\epsilon$ values about one order of
magnitude larger than in UCB: $\epsilon \le 10^{-6}$.

However, it is important to remember that our high energy physics
colleagues are by no means sure what form the high density equation of
state should take.  It may well be that neutron stars contain exotic
solid cores, or that some or all compact objects are not neutron
stars at all but are so-called \emph{strange stars}, consisting of a
mixture of up, down and strange quarks, not arranged into nucleons.

The problem of calculating a maximum mountain size from an exotic star
was examined by Owen (2005), who considered two possibilities.  First
he examined the case of solid strange stars, using a shear modulus
proposed by Xu (2003) to explain LMXB quasi-periodic oscillations as
torsional oscillations of a fully solid star.  The consequent mountain
was found to be limited in size to $\epsilon \le 2 \times 10^{-4}$,
orders of magnitude larger than for neutron stars.  Owen then
considered the case of a partly baryonic, partly strange star
(Glendenning 1992), with a gradual phase transition between the two.
The maximum mountain in this case was somewhat smaller, but still
larger than the neutron star estimates: $\epsilon \le 5 \times
10^{-6}$.

A different class of exotic compact object was then studied by Haskell
et al. (2007a), who made use of recently developed crystalline color
superconducting quark core models (Mannarelli et al 2007).  The shear
modulus of such matter is sensitive to somewhat uncertain QCD
parameters, including the density at which the transition form such an
exotic state to normal baryonic matter occurs.  Taking optimistic
(from the point of view of gravitational wave emission) values can
lead to $\epsilon \le 10^{-3}$, about an order of magnitude higher
than for the solid quark stars considered by Owen.  So, current
uncertainties in the high density equation of state do allow for some
very large asymmetries indeed.

Before concluding, a few brief remarks on the status of
magnetically-supported mountains.  Here there is a significant
uncertainty that must be borne in mind: It is not at all clear how the
strength of the \emph{internal} magnetic field (which is mainly
responsible for the quadrupole generation) is related to the
\emph{external} magnetic field (which is the potentially measurable
one, e.g. from pulsar spin-down).  In particular, it is not clear how
the internal field is arranged, as this depends upon the
superfluid/superconducting nature of the core, and also on the star's
equation of state (Haskell et al. 2007b).

In the absence of superconductivity, the internal magnetic field is
expected to be uniformly distributed in the core.  The ellipticity
produced by a field of strength $B$ would then produce an ellipticity
that can be estimated by taking the ratio of magnetic to gravitational
binding energies, giving $\epsilon \approx 10^{-12} (B/10^{12}{\, \rm
  G})^2$.  Superconductivity complicates this picture: If the interior
contains a type II proton superconductor, the field is not distributed
uniformly, but is confined to a large number of $10^{15} G$ flux tubes
(see Cutler 2002 and references therein).  This increases the
resulting deformation by a factor of $(10^{15} G)/B$.  More
exotically, if the interior protons form a type I superconductor, the
magnetic field is excluded from the core completely, being forced into
a thin shell at the base of the crust.  As described by Bonazzola \&
Gourgoulhon (1996), the resulting ellipticity diverges as the
thickness of this shell goes to zero, making placing an upper limit
problematic.  Finally, if a star is accreting from a companion,
funneling of the accreted material at the poles might build up a
magnetic mountain, a possibility studied recently by Payne and
Melatos (2006).  Certainly, there are many possibilities when it comes
to producing magnetic deformations, and it is not yet clear if these
mechanisms are competitive with more conventional stressed elastic
crust scenarios.

To sum up, if compact objects really are neutron stars, then crustal
strains allow $\epsilon \le 10^{-6}$.  This is comfortably large
enough to allow for the LMXBs to be spin-limited by gravitational
weave emission, and for the spin-down of the millisecond pulsars to
have a significant gravitational wave component.  This is also just
large enough to be of interest for current gravitational wave
observations: The S3/S4 results recently posed by the LIGO Scientific
Collaboration (LSC) give an upper bound of $\epsilon_{\rm max} = 7.1
\times 10^{-7}$ for PSR J2124-3358, so this non-detection has already
told us something non-trivial: this neutron star at least is not
maximally strained.

The larger possible mountains that exotic compact objects can provide
allow us to make more use of the LSC upper limits.  The S3/S4 results
contain many upper limits of less than $10^{-4}$, telling us that
these objects are not maximally strained strange stars.  These results
also rule our some of the more extreme magnetic field configurations
of Bonazzola \& Gourgoulhon (1996).

The sensitivity of these gravitational wave searches scales with the
noise level in the detector and as the square root of the duration of
the (coherent) observation.  This means that as the detectors are
improved and longer stretches of data analyzed more and more stars
will become potentially detectable, even within the canonical neutron
star scenario.  Looking ahead, a year's worth of data from Advanced
LIGO would provide an upper limit of $\epsilon_{\rm max} \approx
10^{-8}$ for PSR J2124-3358, a level where one no longer feels one has
to be optimistic to make a detection.

We will end this summary by posing two questions, both of importance
for astrophysical interpretation.  Firstly, suppose a positive
gravitational wave detection was made.  What would we learn?  Could we
distinguish between, say, a canonical neutron star carrying a large
strain, or a less highly strained strange star?  Secondly, if even
Advanced LIGO detects nothing, what can we conclude?  Would this rule
out crystalline color superconducting quark cores, or is Nature
capable of producing such phases with ellipticities substantially
below their theoretical maxima?  Clearly, despite the progress of
recent years, there are still plenty of important theoretical issues
to be examined if we are to extract the maximum information from 
the search for
gravitational waves from neutron stars.

{\bf References}

B. Abbott et al, {\em Upper limits on gravitational wave emission from
  78 radio pulsars}, preprint 
\htmladdnormallink{gr-qc/0702039}{http://arxiv.org/abs/gr-qc/0702039}
to appear in Phys.
Rev. D (2007)

L. Bildsten, Ap. J. {\bf 501} L89 (1998)

S. Bonazzola, E. Gourgoulhon, Astron. \& Astrophys. {\bf 312} 675 (1996)

C. Cutler, Phys. Rev. D {\bf 66} 084025 (2002)

N. K. Glendenning, Phys. Rev. D {\bf 46} 1274 (1992)

B. Haskell, N. Andersson, D. I. Jones, L. Samuelsson, Submitted to
Phys. Rev. Lett. (2007a)

B. Haskell, Samuelsson, K. Glampedakis, N. Andersson, Submitted to
MNRAS (2007b)

B. Haskell, D.I. Jones, N. Andersson, MNRAS {\bf 373} 1423 (2006)

M. Mannarelli, K. Rajagopal, R. Sharma, {\em The rigidity of
  crystalline color superconducting quark matter}, preprint
\htmladdnormallink{hep-ph/0702021}{http://arxiv.org/abs/hep-ph/0702021}

B.J. Owen, Phys. Rev. Lett. {\bf 95} 211101 (2005)

J. Papaloizou \& J. E. Pringle, MNRAS {\bf 184} 501 (1978)

D. J. B. Payne, A. Melatos, Ap. J. {\bf 641} 471 (2006)

G. Ushomirsky, C. Cutler, L. Bildsten, MNRAS {\bf 319} 902 (2000)

R. V. Wagoner, Ap. J. {\bf 278} 345 (1984)

R. X. Xu, Ap. J. {\bf 596} L59 (2003)
\vfill\eject
\section*{\centerline{GR18/Amaldi 7  in Sydney 2007}}
\addtocontents{toc}{\protect\medskip}
\addtocontents{toc}{\bf Conference reports:}
\addcontentsline{toc}{subsubsection}{\it
GR18/Amaldi 2007 in Sydney, by Jorge Pullin}
\begin{center}
Jorge Pullin, Louisiana State University\\
\mbox{\ }
\htmladdnormallink{pullin-at-lsu.edu}
{mailto:pullin@lsu.edu}
\end{center}
\parindent=0pt
\parskip=5pt

The 18th International Conference on General Relativity and 
Gravitation (GR18) and the 7th Edoardo Amaldi Conference on
Gravitational Waves were held concurrently in Sydney, Australia,
July 8-14 2007.

Over 600 scientists converged on the Sydney Convention and Exhibition
Center at spectacular Darling Harbour.  There were 15
plenary talks and 55 parallel sessions. A typical day had one
``Amaldi'' and two ``GR'' plenary talks and there were five
``GR'' and only one ``Amaldi'' parallel sessions in the afternoon. 
The format was a bit of a departure from the tradition of Amaldis,
where in the past there was no division between plenaries and
parallels. 

During the conference the Committee of the International Society of
General Relativity met. Among other topics, the results for the
election of the president of the society were announced, Abhay
Ashtekar was elected. The next venue for the GR conference was also
selected, the responsibility going to a proposal from a group of
Mexican institutions to hold the conference at the Banamex conference
center at Mexico City. The Gravitational Waves International Committee
(GWIC) also met before the conference started, and among other things
decided the venue for the next Amaldi Conference, choosing Columbia
University in New York for 2009 among several contenders.

The Basilis Xanthopoulos prize was presented jointly to Martin
Bojowald (PennState) and Thomas Thiemann (Albert Einstein Institute)
for their seminal contributions in loop quantum gravity. The GWIC thesis
prize was awarded to Yoichi Aso (University of Tokyo),
his thesis was on ``Active Vibration Isolation for a Laser 
Interferometric Gravitational Wave Detector 
using a Suspension Point Interferometer''.

A conference as large as this one is impossible to summarize
comprehensively at any level of detail. The plenary program had talks
by Stan Whitcomb (LIGO) on ground based gravitational wave detectors, Laurent
Freidel (Perimeter Institute) on spin foam models of the dynamics of quantum
space-time, Steve McMillan
(Drexel) on gravitational dynamics of large stellar systems, Badri
Krishnan (Albert Einstein Institute) on quasi-local black hole  horizons, Bernd
Br\"ugmann (Jena) on numerical relativity, Daniel Eisenstein (Arizona) on
observing dark energy, Peter Schneider (Bonn) on cosmological probes of 
gravitational lensing, Renate
Loll (Utrecht) on the emergence of space-time in 
dynamical triangulation quantum gravity
Francis Everitt on Gravity Probe B and STEP, Hans Ringstrom (Stockholm)
on some rigorous results on cosmic censorship, Jonathan Feng (Irvine)
on collider physics and cosmology, Daniel Shaddock (JPL) on LISA,
Maria Alessandra Papa (Albert Einstein Institute) on gravitational
wave astronomy from ground and space,
Robert Myers (Perimeter Institute) on the quark soup (``al dente'') at RHIC and
its connection to gravity via string theory, Ralf Sch\"utzold on
possible analogue gravity experiments involving horizons. 
There were lectures to the general public
in the evenings by Kip Thorne (``The warped side of the universe:
from the big bang to black holes and gravitational waves'') 
and Roger Penrose (``What happened before the big bang?''). 

As for a personal point of view, I liked the summaries about the
meeting that mathematician Marni Dee Sheppeard wrote for her
blog. Go to,

\htmladdnormallink{http://kea-monad.blogspot.com/2007\_07\_01\_archive.html}
{http://kea-monad.blogspot.com/2007_07_01_archive.html}

and scroll down a few screens. 

This was the first joint GR/Amaldi.  From comments I gathered people
seemed to like the idea of a joint meeting.  The ``traditional GR
crowd'' liked the more extensive exposure to the emerging area of
gravitational wave detection that the Amaldi provided, and the
``traditional Amaldi'' people welcomed the more broad choice of
plenary talks and the opportunity to venture into parallel sessions of
interest to gravitational wave detection (e.g. numerical relativity)
that may not be well represented in a traditional Amaldi. I believe
there is enthusiasm among the interested parties to repeat the
experiment in the future. The GR's occur every three years and the
Amaldis every two, so perhaps another joint meeting could be possible
at the time of the GR after the next one.

\vfill\eject
\section*{\centerline
{Synergy in Singularities?}}
\addtocontents{toc}{\protect\medskip}
\addcontentsline{toc}{subsubsection}{\it
Synergy in Singularities?
, by Don Marlof}
\parskip=3pt
\begin{center}
Don Marolf, University of California, Santa Barbara 
\htmladdnormallink{marolf-at-physics.ucsb.edu}
{mailto:marolf@physics.ucsb.edu}
\end{center}

We relativists love to talk about singularities.  
Schwarzschild black holes and FLRW cosmologies provide our favorite
examples.  The terms ``geodesic incompleteness,'' ``curvature blow-up,'' and ``quantum resolution'' roll easily off of our tongues.   It will therefore come as no surprise that the January, 2007 mini-program
on ``The Quantum Nature of Spacetime Singularities'' at UCSB's KITP was a scene of provocative  discussion and intense interaction, laced with much speculation about results which are soon to come.

Singularities 2007 was a melting pot type of program, which brought together researchers working on the classical characterization of singularities, loop quantum gravity, and string theory.  Mixing such cultures can be difficult, but also rewarding as each community benefits from seeing their work through the eyes of others.  I am sure that the organizers (M. Bojowald. R. H. Brandenberger, G. T. Horowitz, and H.Liu) were gratified by the time and effort put forth by each community in learning to better understand the other approaches and in responding to questions which probed their own work from new directions.

The interactive cross-cultural atmosphere led to a very interesting series of talks and discussions which are available  (audio, video and slides) on-line at  

\htmladdnormallink{http://online.kitp.ucsb.edu/online/singular\_m07/}
{http://online.kitp.ucsb.edu/online/singular\_m07/}

\noindent
Since a broad set of readers may be interested in the introduction and overview talks, I briefly review such talks below.  (For experts,  I also recommend the more technical or specific talks not listed here. Check out the complete list on-line!)    

The reader will not be surprised that seminars focused on areas of recent progress in each field.  These include: 
\begin{quote}
 i )  The characterization of classical singularities. \newline
 Here the talks by Berger (1/09/07) and Garfinkle (1/10/07) provide a good introduction.  
 
 ii) The `bounce' behavior seen in loop quantum cosmology.  \newline
 Here I can recommend the overview talks by Ashtekar (1/12/07) and Bojowald (1/24/07), while the series of talks (1/17/07 and 1/18/07) by Thiemann fills in background on loop gravity methods and discusses how singularities might be approached in full 3+1 loop gravity.   Pullin (1/19/ 07) also presented a somewhat different approach to spherically symmetric black holes.
 
 iii)  Stringy approaches to singularities. \newline
 Several seminars introduced how AdS/CFT can be used to probe black hole singularities (Shenkar (1/11/07) and Liu (1/23/07)) and cosmological singularities (Hertog (1/18/07)), while Silverstein (1/16/07) introduced some ideas from tachyon condensation.  
\end{quote} 
Below, I will reference some of these talks by the speaker's name.

Many talks and discussions revolved around fundamental questions:  {\it Are} singularities fully resolved by a quantum theory of gravity?    If so, what form does this resolution take?  If not, what would it mean that physics remains singular inside black holes or at the big bang?  And, what precisely do we mean by a singularity anyway?

It seems that many of the answers are ``it depends."  First, it remains a logical possibility that either black hole or big bang singularities really do represent the end of meaningful dynamics and in this sense remain ``singular'' even in a fully complete theory.  There is some evidence that this happens for at least some singularities in string theory.  For example, in AdS/CFT when the boundary theory itself is singular (see talks by Shenkar and Hertog), and perhaps in tachyon-condensation associated with the big bang (Silverstein).  While black hole evaporation in string theory generically appears to be a much smoother process (Shenkar), exceptions may arise in the black hole interior (Silverstein).

The above evidence suggests that, within string theory,  evolution starting from a large classical regime and proceeding through a collapsing cosmological singularity does {\it not} result in another a large spacetime region that is even moderately smooth and classical.    
However, precisely the opposite situation holds in loop quantum gravity, where the current signs from simple models (Ashtekar, Bojowald) are that that both black hole evaporation and big bang singularites may be quite `smooth' and result in new quasi-classical regions of spacetime.    While the story in full 3+1 loop gravity remains unclear (Bojowald, Thiemann), 
many simple models feature a `bounce'  at the Planck density.  That is to say that collapsing dimensions of space tend to slow, stop, and then begin to expand once the matter fields reach Planck density and the spacetime reaches Planck curvature.  The resulting expansion then produces another large region of spacetime.   Since the reader will recall that classical de Sitter space also features `bounce' behavior, it will come as no surprise that this loop-gravity bounce also shows features of inflation, raising the tantalizing possibility that loop quantum cosmology might replace standard inflation as a theory of structure formation.  

Further results are eagerly awaited on all fronts.  However, in the end it may be the meta-questions that are of greatest interest.  Do loop gravity, string theory, and other approaches predict radically different consequences as current indications suggest?  If so, then even with the paucity of experimental data which directly probes the quantum gravity regime, experiment and observation may nevertheless be able to rule out certain models.  On the other hand, suppose that current expectations are wrong,  and that all approaches eventually agree on at least the qualitative character of their predictions.  Such a result would require an approach-independent explanation.  As an optimistic theorist, I would expect its discovery to reveal some new and deep truth about the fundamental nature of quantum gravity.
\vfill\eject

\section*{\centerline
{Institute for Gravitation and the Cosmos}}
\addtocontents{toc}{\protect\medskip}
\addcontentsline{toc}{subsubsection}{\it
Gravitation and the Cosmos
, by Derek Fox and Parampreet Singh}
\parskip=3pt
\begin{center}
Derek Fox, Pennsylvania State University 
\htmladdnormallink{dfox-at-astro.psu.edu}
{mailto:dfox@astro.psu.edu}
\end{center}
\begin{center}
Parampreet Singh, Perimeter Institute for Theoretical Physics 
\htmladdnormallink{psingh-at-perimeterinstitute.ca}
{mailto:psingh@perimeterinstitute.ca}
\end{center}

For the past 14 years, Penn State has had first a Center and then
an Institute for Gravitational Physics and Geometry. Researchers
at the Institute came from physics, mathematics and astronomy and
astrophysics backgrounds and have contributed actively to the
interface of these disciplines. To create bridges between the
theoretical work pursued at the institute and the rich science
resulting from state of the art  observations related to the
highest energy phenomena in the universe, the scope of the
institute has now been enlarged to encompass particle
astrophysics. The greater institute is called the Institute for
Gravitation and the Cosmos and was inaugurated through a 3-day
international conference, August 9-11, which brought over 130
participants to the University Park campus of Penn State.

One of the highlights of the conference was the Forum on Science
and Society, a plenary session, where the organizers explained the
vision for the new institute. It will have three Centers: a Center
for Fundamental Theory, a Center for Gravitational Wave Physics
and a Center for Particle Astrophysics. By fostering an active
exchange of ideas between these Centers, the Institute will strive
to create new opportunities and open novel directions at the
interface of these mature fields. The organizers gave a few
examples to illustrate this vision. They will seek to bring
together experts in loop quantum gravity, string theory and
cosmology to address fundamental physical issues in the hope that
this multi-pronged approach will reveal new avenues which
transcend individual areas. Another example came from the fact
that the new Institute has research groups dedicated to exploring
the universe using all four fundamental forces of Nature: strong
interactions through the Pierre Auger cosmic ray project, weak
interactions through the IceCube neutrino experiment,
electromagnetic through the {\it Swift} gamma ray burst explorer mission
with its headquarters at  Penn State, and gravitational
interactions through the LIGO  detector.

There were nine technical plenary lectures by international
leaders in various fields covered by the Institute. Each morning
featured three of these, each talk covering a key area in one of
the three Centers. The speakers presented excellent overviews
which could be appreciated by the diverse audience. The
juxtaposition of talks from very different areas brought out not
only the intellectual breadth but also common themes underlying
apparently distinct areas.

{\it Joe Polchinski} (UC, Santa Barbara) spoke about the black
hole information loss issue; {\it Slava Mukhanov} (Munich)
summarized features of the inflationary scenario using a
model-independent paradigm; and {\it Roger Penrose} (Oxford and
Penn State) summarized ideas on the arrow of time, the nature of
the big bang and the necessity of information loss that he has
developed over the last three years. {\it Frans Pretorius}
(Princeton) gave an overview of the history and the current status
of the binary black hole problem; {\it Cliff Will} (Washington
University) described the unreasonable success of post-Newtonian
methods and {\it Karsten Danzmann} (AEI, Hannover) gave a survey
of the proposed multi-messenger gravitational wave astronomy. {\it
Jim Cronin} (Chicago) provided a historical overview of cosmic-ray
astronomy with updates including the `break' in the spectrum near
$E = 10^{19.5}$ eV; {\it Thomas Gaisser} (Bartol) discussed the
connections between cosmic ray and neutrino observations through
upcoming experiments such as IceCube and the Pierre Auger
projects; and {\it Trevor Weekes} (Whipple Observatory and
Smithsonian Center for Astrophysics) reviewed the history of
ground-based TeV gamma-ray astronomy and provided updates
focusing  on results from HESS and VERITAS.

These overviews brought out the spectacular progress that has
occurred on both theoretical and observational fronts in recent
years. The fact that several of the observational missions are
likely to have major new results in the coming years brought out
the excitement of the current state of fundamental physics.

The afternoons featured 13 parallel sessions; two on cosmology,
four on approaches to quantum gravity, two on numerical
relativity, two on gravitational waves and one each on
observational issues in particle astrophysics, the origins of high
energy particles, and astro-particle physics beyond the standard
model. Typically, the lead talk was an invited contribution and
set the stage for the session. In Cosmology, the lead speaker was
{\it Mark Trodden} (Syracuse) who began by describing the current
challenges and then gave a brief summary of the current ideas from
particle physics as well gravity communities; in quantum gravity,
{\it Laurent Friedel} (Perimeter) described how effective,
low-energy theories arise from spin foam models in loop quantum
gravity and how they naturally make contact with non-commutative
field theories; and in numerical relativity,  {\it Manuel Tiglio}
(LSU) described new approaches to binary black hole evolutions. In
particle astrophysics {\it David Seckel} (Bartol) discussed a
variety of approaches that will be used to search for ``GZK''
neutrinos, emphasizing the advantages of radio-Cerenkov
techniques; {\it Kaixuan Ni} (Yale) described cryogenic particle
dark matter searches and constraints on WIMP  models; and in the
session on gravitational waves, {\it Ben Owen} (Penn State)
described LIGO's diverse search methods for periodic signals  from
neutron stars and other Galactic sources, including the
`Einstein@Home' project.

There were 45 contributed talks in various sessions  featuring
many interesting results. For example, {\it John Carrasco} (UCLA)
described the intriguing results indicating that $N=8$
supergravity may after all be finite in 4 space-time dimensions;
{\it Kevin Vandersloot} (Portsmouth) explained how quantum
geometry effects in loop quantum cosmology manage to resolve the
big bang singularity and yet quickly fade away to ensure agreement
with classical general relativity; {\it Yi Pan} (Maryland)
explained how the effective one body approximation and numerical
relativity can be used, hand in hand, to develop better templates
for LIGO; {\it Emmanouela Rantsiou} (Northwestern) discussed
numerical simulations of black hole-neutron star mergers with
examples of the disrupted neutron star forming an extended tidal
tail; and {\it Seon-Hee Seo} (Penn State) described various data
analysis techniques used in IceCube to distinguish rare tau
neutrino interactions. 

The Forum on Science and Society also featured short invited talks
by {\it Roger Penrose} and three leaders from industry who are
deeply involved in issues at the interface of science and society
---{\it Dr. Edward Frymoyer}, a leader in the Fibre Channel
Technology; {\it Mr. Christopher Liedel}, Senior Vice President
and Chief Financial Officer of the National Geographic Society,
Chairman of the National Philanthropic Trust and the primary force
behind creation of the FQXi, a community dedicated to fundamental
issues in cosmology and physics; and  {\it Mr. Duane Valz}, Vice
President and Associate General Counsel for Global Patent Strategy
at Yahoo! Inc. Each of these speakers emphasized the importance of
strong and mutually beneficial relationship between the public and
the scientific community. They stressed the significance of the role
scientists play in society, the way they can promote a
scientific temperament in society, how they can create public
awareness of the fascinating research they are involved in, and
how in turn the public can contribute to the sustenance of research.
In particular, {\it Chris Liedel} explained the way National
Geographic works closely with scientists at various universities
and institutions to disseminate research to the general public and has
funded fundamental research by using resources gained from its
popular magazines and documentaries.

The conference banquet was a warm, festive occasion with a
Mediterranean dinner. Several in the audience gave personal
accounts of how the Penn State Institute has provided them with
intellectual stimulation. In particular, the junior faculty in the
Institute described the unique atmosphere that fosters their
interdisciplinary research. There were also some suggestions as to
what the Institute could do to enhance the public awareness of
forefront science.

Many participants commented that the conference provided an unique
perspective on a wide variety of forefront issues and brought out
a surprising number of inter-relations.

\vfill\eject
\section*{\centerline
{Post-Newtonian Theory
and Numerical
Relativity 
}}
\addtocontents{toc}{\protect\medskip}
\addcontentsline{toc}{subsubsection}{\it
NumRel meets PN
, by Buonanno et al}
\parskip=3pt
\begin{center}
Alessandra Buonanno, University of Maryland
\htmladdnormallink{buonanno-at-umd.edu}
{mailto:buonanno@umd.edu}
\end{center}
\begin{center}
Gregory Cook, Wake Forest University 
\htmladdnormallink{ cookgb-at-wfu.edu}
{mailto: cookgb@wfu.edu}
\end{center}
\begin{center}
Sam Finn, Penn State University  
\htmladdnormallink{lsfinn-at-psu.edu}
{mailto:lsfinn@psu.edu}
\end{center}
\begin{center}
Pablo Laguna, Penn State University
\htmladdnormallink{pablo-at-astro.psu.edu}
{mailto:pablo@astro.psu.edu}
\end{center}

The Washington University Gravity Group 
hosted a dynamic and exciting meeting in St. Louis 
entitled ``Numerical Relativity meets 3PN: A Workshop'', from
February 8-11, 2007.  In
attendance were many of the leading researchers in the fields of
numerical relativity, post-Newtonian theory, and gravitational-wave
data analysis.  The purpose of the meeting was to bring these
researchers together in an effort to stimulate progress at the
important interface between these fields.  The workshop was organized
with a set of invited talks each morning, designed to probe the
central questions and problems with ample time for discussions.  Each
afternoon started with a small number of short contributed talks
followed by an open ended working/discussion session focused on a
different topic each afternoon.  The workshop program can be found at
\htmladdnormallink{http://nrm3pn.wustl.edu/}
{http://nrm3pn.wustl.edu/}.
Below, we summarize the highlights of
the meeting and attempt to give proper credit to individuals and
the groups they represent.  We apologize for any errors or omission,
and for the fact that we cannot mention everyone who contributed
to this very successful meeting.

During the workshop all of the major research groups performing
simulations of black-hole binary collisions presented their most
recent results [Pretorius (Alberta/Princeton); 
Lousto (UT Brownsville); Baker (NASA Goddard);
Scheel (Cornell/Caltech); Hannam, Husa (Jena/AEI); Laguna (Penn State)],
along with results from groups simulating NS-NS collisions [Suen (WU)] and
NS-BH collisions [Faber(UIUC)]). But the focus was on the BH-BH
inspiral phase which can be compared with PN models.  Prior to the
workshop, various groups had obtained simulations with as many as 8
orbits (16 gravitational-wave cycles) prior to merger.  These
simulations showed good agreement with analytic PN results at 3.5PN
order [Baker; Buonanno (UMD)]. 
More specifically, standard PN models (expanded form of the
balance equations) and the effective-one-body (EOB) model at 3.5PN
match well with the inspiral waveforms obtained using the generalized
harmonic and moving puncture codes.  While the initial agreement is
good, longer and more accurate simulations will determine how
significant the dephasing is between the numerical and analytic
predictions.

A significant problem with all of the simulations is that
the initial data
used by most groups place the binary in a ``quasi-circular'' orbit with
an effective eccentricity that modulates both the amplitude and the
phase of the gravitational wave. Scheel 
showed preliminary results from the Cornell/Caltech group
for an inspiral waveform emitted by a
non-spinning equal-mass binary where the eccentricity had been reduced
to $10^{-6}$.  The trajectory of a non-spinning equal-mass black-hole
binary moving along an adiabatic inspiral is unique, and this was the
first time such an evolution had been simulated.  Interestingly,
the evolution was obtained without using PN initial data.

One of the more important aspects of the group discussions concerned
how to quantify the differences between gravitational-wave signals,
whether the differences are between NR and PN waveforms, or between
waveforms created by different codes or with different initial data.
Much of this discussion was guided by the data-analysis community
[Sathyaprakash (Cardiff); Owen, Finn (PSU)].  There 
were also discussions of which kind of
comparisons should be pursued in the future.  Should one compare
trajectories in the same gauge and coordinates, or focus on more
invariant quantities?

While the talks and discussion naturally focused largely on the
inspiral phase, other aspects of black-hole binary collisions were
discussed.  Some of the most surprising and intriguing new results
presented at the workshop concerned the recoil velocity of the final
black hole following binary coalescence [Lousto;
Sperhake (Jena/AEI); Laguna].  The first two numerical
relativity groups showed that for special spin configurations the kick
can be unexpectedly large.  Analytic studies aimed at understanding
how the kick builds up during the inspiral, merger and ringdown
phases were presented.

The ringdown signal following coalescence certainly contains valuable
information and it is possible to extract a few of the quasi-normal
modes of Kerr from these ringdown waveforms.  Results were shown both
for equal-mass [Buonanno] and unequal-mass binary systems [Berti (WU);
Tiglio (LSU)]. 
Most of the presentation and discussion concerned new techniques to improve
the fits.

The transition region connecting the end of the inspiral through the
merger to the ring down was also considered.  Combining an EOB
waveform with a superposition of a fundamental quasi-normal mode plus
two overtones by matching them at the light ring [Buonanno], it was
shown that the ringdown phase could be modeled reasonably well.
However, we lack an analytic description of how the quasi-normal modes
are excited, whether non-linearities and/or mode-mixing are important.
Understanding these issues is needed to improve the analytic matching
of the inspiral to ring down.

There were long discussions on data analysis issues.  Considering the
current numerical results and studies, it seems that simple
modifications of existing PN and resummed-PN template banks, guided by
NR results, should lead to high matching performances with numerical
waveforms and should be used for signal detection.  However, for
parameter estimation, we need to improve the current PN template
banks.  There were also discussions on if and by how much one should
expand the template bank if one thinks that the real waveforms deviate
from PN or NR predictions.

Finally, diagnostics of NR results through PN tools [Cook (Wake Forest); 
Will (WU)] 
and initial-data issues were discussed.  New approaches for
implementing PN initial data were presented [Blanchet (IAP Paris); 
Tichy (FL Atlantic)].
A central goal of these approaches is to incorporate a physically
realistic gravitational-wave signal into the initial data.  It is
hoped that this will help one to understand and perhaps eliminate the
initial burst of unphysical radiation seen in current evolutions.  In
spite of the initial burst of unphysical radiation, the current
families of numerical initial data are proving to be very effective.
Considering that one can reduce the eccentricity and get the unique
adiabatic inspiral, there were discussions as to whether or not one needs
to improve the initial data.  There were no definite conclusions, but
it seems that the current data may be sufficient for binaries on
circular orbits.  However, more work needs to be done with spinning
and precessing binaries, and eccentric binary systems.

The meeting ended with a panel discussion. The moderator was Sam Finn,
with Alessandra Buonanno, Greg Cook and Pablo Laguna serving as
panelists.  Finn started the session with introductory remarks about
the theme of the NRm3PN meeting, namely how NR meets 3PN in
understanding dynamics, nonlinear theory, applications to data
analysis, and astrophysics. Next, each of the panelists was asked to
recall the most interesting, unexpected, exciting or novel thing
learned at the workshop.  Buonanno responded that the work presenting
the comparison of PN and NR waveforms was very encouraging and that the work
presented by the Caltech/Cornell group, which succeeded in reducing the
eccentricity to negligible values, showed, for the first time, the true,
unique inspiral for an equal-mass binary on a quasi-circular orbit. 
Both Cook
and Laguna pointed to the results presented by several groups on kicks
as being extremely interesting, with Laguna pointing out that they
also provide validation of the computational infrastructure. The panel
session ended with Finn posing several forward looking questions: 1)
What is the single-most important question that needs to be resolved
for NR to become useful for data analysis? Buonanno's answer was to generate
longer and more accurate waveforms for spinning, precessing binary systems;
to understand more in detail how the ringdown
modes are excited at the end of the inspiral phase; 2) What is the most
tantalizing indication of unexpected,
novel or surprising behavior associated with non-linear gravity?  Cook
responded that in his opinion it would be the lack of a clearly
distinguishable dynamical plunge prior to merger;  3) What are the
questions arising from the recent results on gravitational recoil?
Laguna pointed out that current results have only probed a very small
fraction of the parameter space. The time is right for a systematic
collaborative exploration involving several groups.

The next installment of ``NRm3PN'' will be June 12 - 14, 2008 at
the University of Jena, as part of a conference recognizing Gerhard
Sch\"afer's 60 birthday.

\vfill\eject
\section*{\centerline
{Saul Teukolsky Birthday Symposium
}}
\addtocontents{toc}{\protect\medskip}
\addcontentsline{toc}{subsubsection}{\it
Saul Fest
, by Greg Cook}
\parskip=3pt
\begin{center}
Gregory B. Cook, Wake Forest University 
\htmladdnormallink{cookgb@wfu.edu}
{mailto:cookgb@wfu.edu}
\end{center}

On June 2, 2007, immediately following the 10th Eastern Gravity
Meeting, a one-day celebratory symposium and banquet was held at
Cornell University to honor Saul Teukolsky's upcoming 60th birthday.
To no one's surprise, the event drew a large crowd of Saul's friends,
family, students, and colleagues.

The Symposium's talks covered a range of topics from Saul's broad
interests in gravitation and relativistic astrophysics.  The morning
session of talks included three former post-docs from Saul's group:
Dierdre Shoemaker(Penn State), Monica Colpi(Milano), and Sam Finn(Penn
State), along with Scott Hughes(MIT) who began his career in physics
as a member of Saul's undergraduate scientific visualization team.

Dierdre Shoemaker's talk on ``Binary Black Hole Mergers'' provided an
overview covering both the history of numerical simulations of binary
black hole collisions and the dramatic advances that have taken place
over the last two years in that field.  Paying homage to the Teukolsky
equation, Scott Hughes give a talk on ``Perturbation Theory and Binary
Evolution'', reminding everyone of the importance of the Teukolsky
equation and discussing recent work on its application to the
extreme-mass-ratio inspiral problem.  Monica Colpi gave a fascinating
talk on ``Light and Gravitational Waves from Massive Black Hole
Binaries'', focusing on astrophysical investigations associated with
black-hole binaries.  Finally, Sam Finn talked on ``Gravitational Waves
and Their Detection'', giving an informative overview of the
status of, and future plans for, the current generation of
gravitational-wave observatories.

Following a catered lunch, the afternoon session of talks drew on
speakers Saul had met as a graduate student at Caltech, during the 
``golden age of black-hole physics''.  The featured speakers were Kip
Thorne(Caltech), Saul's PhD adviser, and two fellow Caltech graduate
students, Richard Price(UT Brownsville) and Alan Lightman(MIT).

Leading off the afternoon, Richard Price gave an entertaining talk on 
``Radiative Tails and the Teukolsky Equation'', presenting a new perspective 
on the subtle issue of mode coupling in the Teukolsky equation.  Kip 
Thorne presented a charming talk on ``Saul and the Warped Side of the
Universe''.  Outlining Saul's journey to become a physicist from his
youth in South Africa through his time at Caltech, Kip painted a warm
picture of Saul as a brilliant student who struggled with and
eventually vanquished the problem of how to handle perturbations of
Kerr.  The final presentation of the Symposium was given by
physicist-turned-novelist Alan Lightman.  In a talk titled ``A Sense of
the Mysterious'', he presented his unique perspective on how scientists
and artists often view the world in very different ways.

The day of talks was capped off by a banquet featuring Bill Press(UT
Austin) as the after-dinner speaker.  Another of Saul's fellow Caltech
graduate students, Press presented a thoroughly entertaining talk and
slide show.  In pictures secretly provided by an "anonymous source",
we got to see how Saul really prefers to dress when carrying out
complex calculations at the kitchen table, and we learned of his
surprising connection to Zaphod Beeblebrox.

Adding to the after-dinner entertainment, ``Bernie and the Gravitones''
presented ``Don't $Psi_4$ Me, Saul Teukolsky'' sung to the tune of ``Don't
Cry for Me, Argentina''.  With Bernie Shutz unfortunately unable to
attend (although he was present in spirit via his picture on an
easel) Cliff Will(Wash. U) bravely led Richard Price, Kip Thorne,
Sandor Kovacs(Wash. U), and Eanna Flanagan(Cornell).

Saul Teukolsky is modest man, but if a person is to be measured by his
work and by his friends, then the day's festivities showed Saul
has much to be proud of.

\vfill\eject
\section*{\centerline
{3rd Gulf Coast Gravity Conference
}}
\addtocontents{toc}{\protect\medskip}
\addcontentsline{toc}{subsubsection}{\it
3rd Gulf Coast Gravity Conference
, by Vitor Cardoso}
\parskip=3pt
\begin{center}
Vitor Cardoso, University of Mississippi
\htmladdnormallink{vcardoso-at-phy.olemiss.edu}
{mailto:vcardoso@phy.olemiss.edu}
\end{center}

The Gulf Coast Meetings started in Brownsville two years ago, and as Richard Price reported back then
\cite{Pullin} they were designed to bring together the growing number of relativists in the gulf coast region
of the country.

The 3rd Gulf Coast Gravity meeting took place at the University of Alabama in Huntsville on January 23-24
(2007), and was hosted by Lior Burko. Financial support was generously provided by the Physics Department at
UAH. The meeting would not have been possible without support from Cindi Brasher and Dora Wynn, and assistance
by the UAH Society of Physics Students. Huntsville (``the rocket city'') features the NASA-Marshall Space
Flight Center, which makes for a good visit after a meal at any  one of the several very nice restaurants
scattered through downtown or University Avenue.

The meeting consisted of short, 12 min talks with particular emphasis on student presentations (there is an APS
sponsored GGR Prize for the best student talk) and quick overviews of some recent developments. Important
breakthroughs in numerical relativity have led to a flurry of activity in the field, and most presentations
focused on numerical relativity and on gravitational wave emission.

The conference opened with Michael Watson describing progress in simulating jet formation with particle-in-cell
codes \cite{watson}, using the Kerr geometry as a fixed background. Preliminary results show that a jet {\it
is} formed. Further details, such as Lorentz-factors and jet distribution dependence on the model parameters
will be dealt with in the future. Alan Farrell described recent progress in understanding the numerical
instabilities affecting some of the codes for neutron star structure. Preliminary results for isolated neutron
stars are encouraging, generalization to more complex systems such as binaries is in progress. Peter Diener
made a concise summary \cite{diener} of recent work on kicks from generally spinning binaries. He explained the
importance of having accurate estimates for the kick velocity from black hole mergers, and how crucial these
numbers may be for our understanding of galactic evolution. Enrique Pazos talked about the influence of the
background spacetime on the accuracy of wave extraction \cite{Pazos}, and particularly about the effect of
extraction radius on the measured waveforms. He exemplified the problem with Gaussian wave pulses scattering
off a Schwarzschild black hole, both in a linear and non-linear analysis. Arunava Roy described how unification
of gravity with electroweak forces may be achieved in extra-dimensional scenarios. In this eventuality, black
holes may be created in particle accelerators and cosmic rays, and their signatures can be studied in detail
using GROKE \cite{groke}, a Monte-Carlo generator for black hole production. Miguel Megevand implemented a
theoretical model for a scalar field to try to mimic accreting systems. A model with a non-trivial
position-dependent potential displays some of the expected features of accreting systems.

The first talk of the afternoon session began with Richard Price exposing some of the contradictions in the
literature on late-time tails in the Kerr geometry. He proceeded by showing how mode coupling can be understood
already at the linearized level as a second order effect in the rotation parameter, and how this may reconcile
the existing contradictory results. Following on Richard's talk, Carl Blaksley described his recent results on
tails in the vicinities of extreme charged black holes, and how they differ from previous predictions. The
non-triviality of these results follows from the special behavior of the effective potential for wave
propagation near the horizon. Manish Jadhav reported on progress to understand how well LISA performs at doing
black hole spectroscopy. In particular he's focusing on computing signal-to-noise ratios as a function of
source location in the sky, using a Markov Chain Monte Carlo generator. Yasushi Mino reviewed the present
status of radiation reaction and self-force on curved backgrounds \cite{Mino}, of particular interest for
extreme-mass ratios inspirals. He explained how to resolve some discrepancies and conceptual obstacles in
understanding the self-force, and how to proceed in the quest for a logical basis for self-force calculations.
In the final talk of the first day, Scott Hawley reported on his recent results \cite{Hawley} concerning
initial data for spinning black holes and their interpretation via classical results. An enlarged parameter
survey, pushing the separation of the holes to larger values, will possibly clarify some of the remaining
issues.

The first day closed with the award of the prize for the best student talk to Enrique Pazos, for his
presentation ``The effects of the background geometry on the extracted waveforms''. Congratulations Enrique!

The second day of the meeting opened with Marco Cavagli\`a showing how uniform motion generates gravitational
radiation in braneworld scenarios. This effect may lead to potentially observable consequences
\cite{Cardoso:2006nz}, such as energy loss by cosmic rays and a stochastic background of gravitational
radiation. Pedro Marronetti continued Diener's kick talk to show us that results displaying massive kicks with
optimally aligned spin are solid ones \cite{Pedro}, in the sense that deviations from initial data producing
the largest kicks still produces large kicks. Pedro stressed the need to have an understanding of the
astrophysically relevant spin distributions in binaries; this will eventually determine whether or not large
kick velocities are common. The good news is that the computational power to perform some of these calculations
is modest. Yes, you can try this at home! I presented recent results on unequal-mass mergers of black hole
binaries \cite{Berti}, and discussed the nature of the waveforms. I tried to convince the audience that most of
the waveform is well understood: the PN approximation describes very well the inspiral almost all the way to
the formation of a common apparent horizon, and ringdown describes extremely well the last part. The results
show that black hole spectroscopy is possible and also that we are ready to start building realistic templates
for the detection of gravitational waves from binary inspiral. Chris Beetle elaborated on an interesting
alternative to Post-Minkowskian expansions, the Periodic Standing-Wave Approximation \cite{Beetle},
specifically designed to handle periodic systems such as appropriately spaced binaries. He explained the
similarities between these two methods and the advantage of having alternatives which can be implemented
numerically to all orders. Our host Lior Burko
presented results for a time-domain implementation of Teukolsky
equation with a delta-like source term \cite{Lior}. He emphasized the need to have credible alternatives to the
well-established frequency-domain methods, and presented in detail 
results for particles in eccentric and
parabolic orbits around rotating black holes. Brett Bolen showed us how to obtain consistent cosmologies in
extra-dimensional scenarios with Gauss-Bonnet terms \cite{Andrew}, where the size of the extra dimensions is a
time-varying function. In this model, corrections to the usual FRW cosmology are consistent with a dark energy
equation of state. Closing the meeting, Ruslan Vaulin explained how to use the trace anomaly to compute
consistently the stress energy tensor of quantum fields in black hole spacetimes \cite{Mottola}. Ruslan
presented several instances of this computation, showing how to obtain non-divergent renormalized stress-energy
tensors throughout the spacetime, and how well this method compares to other alternative methods.

The 4th Gulf Coast Gravity Meeting will be held at the University of Mississippi in Oxford, MS. We hope to see
you all there!


\bibliography{paper}

\end{document}